\documentclass[12pt,fleqn]{article}
\usepackage{times}
\usepackage{fourier} 
\usepackage{graphicx}
\usepackage{wrapfig}
\usepackage{amssymb}
\usepackage{amsmath}
\usepackage{color}
\usepackage{cite}
\usepackage{ulem}

\begin{document}
\baselineskip 0.6cm
\renewcommand{\thefootnote}{\#\arabic{footnote}} 
\setcounter{footnote}{0}
%
\begin{titlepage}
\begin{center}

\begin{flushright}
\end{flushright}


{\Large \bf 
Neutrino masses and gravitational wave background
}

\vskip 1.2cm

{
Takehiko Asaka$^1$
and 
Hisashi Okui$^{2}$
}

\vskip 0.4cm

$^1${\it
\small 
  Department of Physics, Niigata University, Niigata 950-2181, Japan
}

$^2${\it \small
Graduate School of Science and Technology, Niigata University, Niigata 950-2181, Japan}

\vskip 0.2cm

(December 25, 2020)

\vskip 2cm

\begin{abstract}
We consider the Standard Model with three right-handed neutrinos 
to generate tiny neutrino masses by the seesaw mechanism.
Especially, we investigate the case when one right-handed neutrino 
has the suppressed Yukawa coupling constants.
Such a particle has a long lifetime 
and can produce an additional entropy by the decay.
It is then discussed the impact of the entropy production on the gravitational wave background 
originated in the primordial inflation.
We show that the mass and the coupling constants
of the long-lived right-handed neutrino can be probed
by the distortion of the gravitational wave spectrum,
leading to the information of the mass of the lightest active neutrino.
\end{abstract}
\end{center}
\end{titlepage}

\section{Introduction}
Our understanding of neutrinos has been greatly improved from 
the end of the last century.  Various oscillation experiments have 
provided the evidence of neutrino masses.
The observational data are consistent with the 
flavor oscillations by three active neutrinos $\nu_i$ ($i=1,2,3)$
with masses $m_i$.  
However, unknown properties of neutrinos still exist, including
the absolute mass scales,
the Dirac or Majorana nature, and the CP violating phases.
As for the neutrino masses, there are two possible mass orderings. One is the normal ordering (NO) with $\Delta m_{21}^2 = 7.42^{+0.21}_{-0.20} \times
10^{-5}$~eV$^2$ and $\Delta m_{31}^2 = 2.517^{+0.026}_{-0.028} \times
10^{-3}$~eV$^2$, and the other is
the inverted ordering with
$\Delta m_{21}^2 = 7.42^{+0.21}_{-0.20} \times
10^{-5}$~eV$^2$ and $\Delta m_{32}^2 = -2.498^{+0.028}_{-0.028} \times
10^{-3}$~eV$^2$~\cite{Esteban:2020cvm}, which shows that
the absolute values of neutrino masses or
the mass of the lightest active neutrino,
denoted by $m_0$, is undetermined so far.
Note that $m_0=m_1$ or $m_3$ for the NO or IO case.
The sum of neutrino masses is $\sum m_i < 0.12$~eV from 
the cosmological constraints~\cite{Aghanim:2018eyx} (see also Ref.~\cite{Vagnozzi:2017ovm}),
which leads to $m_0 < 0.030$~eV or 0.016~eV 
for the NO or IO case, respectively.

Furthermore, the mechanism for generating the non-zero neutrino masses
is unknown yet.  One of the most attractive ways to explain
the tiny neutrino masses is the seesaw mechanism by right-handed 
neutrinos~\cite{Minkowski:1977sc,Yanagida:1979as,Yanagida:1980xy,Ramond:1979,GellMann:1980vs,Glashow:1979,Mohapatra:1979ia}.
Here we consider the case where the number of right-handed neutrinos is three.%
\footnote{The mass of the lightest active neutrino is $m_0=0$
for the case with two right-handed neutrinos.}
In this case the possible region of $m_0$ is below the above bound 
since there is no reason to select a specific value of $m_0$.%
\footnote{There are, of course, possibilities to determine the 
scale of $m_0$ by introducing 
an additional mechanism to the theory such as
the flavor symmetry.}
Especially, when $m_0 \ll {\cal O}(10^{-3})$~eV, the determination
of $m_0$ by neutrino experiments becomes very hard.

In such a situation one of three right-handed neutrino, say $N_S$, can have
the Yukawa interactions with very suppressed couplings
and become a very long-lived particle, and then an additional entropy
can be produced by the $N_S$ decays and the universe is reheated 
again at late epoch after the reheating of the primordial inflation.
(See, for example, Refs.~\cite{Asaka:2006ek, Ghiglieri:2019kbw}.)
This entropy production dilutes the pre-existing dark matter,
baryon asymmetry, and dangerous long-lived particles in cosmology.

In addition, it modifies the thermal history of the universe 
and the spectrum shape of the primordial gravitational wave (GW) 
background,
which is a good target for the future observations.
This issue has been investigated in Refs.~\cite{Seto:2003kc,Nakayama:2008ip,Nakayama:2008wy,Kuroyanagi:2011fy,Buchmuller:2013lra,Jinno:2013xqa,Jinno:2014qka,Kuroyanagi:2014nba,Kuroyanagi:2014qza,DEramo:2019tit,Blasi:2020wpy,Kuroyanagi:2020sfw}.
It has been shown that the entropy production leads to
the suppression of the GW spectrum at high frequencies, 
from which the reheating temperature $T_R$ of the entropy production
and the rate $\Delta S$ between
the entropy before and after the decay can be probed by the distortion signature
of the GW spectrum.

In this paper we discuss the entropy production by the decays of right-handed neutrino $N_S$ in the seesaw mechanism.  Especially, 
we consider the case when the mass of $N_S$ is heavier than ${\cal O}(1)$~TeV
and $m_0 < {\cal O}(10^{-7})$~eV, and then discuss
the impacts on the primordial GW background spectrum. 
It is then shown that the mass and Yukawa coupling of $N_S$ can be examined 
by the GW spectrum shape, which results in the determination of $m_0$.
Remarkably, the suppression rate of the spectrum is directly related to 
$m_0$. 

The paper is organized as follows.
In the next section, we explain the framework of the analysis
and demonstrate how the decays of right-handed neutrino lead to 
the late time production of an additional entropy.
In section 3, we present the spectrum distortion of the primordial GW 
background
by the entropy production and show what can we learn from it.
The final section is devoted to the conclusions.

\section{Seesaw mechanism and entropy production}
We consider the Standard Model which is extended by three right-handed neutrinos
$\nu_{RI}$ ($I =1,2,3$) with Lagrangian
\begin{align*}
    {\cal L} = {\cal L}_{\rm SM}
    + i \overline \nu_{RI} \gamma^\mu \partial_\mu \nu_{RI}
    - \left(
    F_{\alpha I} \overline{L}_\alpha H \nu_{RI}
    +
    \frac{M_I}{2} \overline \nu_{RI}^c \nu_{RI}
    + h.c.
    \right) \,,
\end{align*}
where the Higgs and left-handed lepton doublets are denoted by
$H$ and $L_\alpha$ ($\alpha = e, \mu, \tau)$, respectively.
$F$ is the Yukawa coupling matrix and $M_I$ are the Majorana masses 
of right-handed neutrinos.  Note that we take the basis in which 
the mass matrices for charged leptons and right-handed neutrinos are diagonal.

We assume the hierarchy between the Dirac masses 
$|[M_D]_{\alpha I}|=|F_{\alpha I}|\langle H \rangle$ and $M_I$ 
for the seesaw mechanism.
The mass matrix of active neutrinos $\nu_i$ ($i=1,2,3)$ is given by
\begin{align}
    [M_\nu]_{\alpha \beta} = - [M_D]_{\alpha I} [M_D]_{\beta I} M_I^{-1} \,,
\end{align}
and the diagonalization of $M_\nu$ gives the neutrino mixing matrix $U$,
called PMNS matrix, as $U^\dagger M_\nu U^\ast=
D_\nu = \mbox{diag}(m_1, m_2, m_3)$
where $m_i$ is the mass for $\nu_i$.
On the other hand, the heavier states $N_I \simeq \nu_{RI}$, called as heavy neutral leptons 
(HNLs), have 
the mixing with left-handed leptons as
$\nu_{L \alpha}
=U_{\alpha i} \nu_i + \Theta_{\alpha I} N_I^c
$ where $\Theta_{\alpha I} = [M_D]_{\alpha I}/M_I$.
Their mass matrix is $D_N = \mbox{diag} (M_1, M_2, M_3)$.
The Yukawa coupling matrix can be parameterized as~\cite{Casas:2001sr}
\begin{align}
    F = \frac{i}{\langle H \rangle} U D_\nu^{1/2} \Omega D_N^{1/2} \,,
\end{align}
where $\Omega$ is the $3 \times 3$ complex orthogonal matrix
($\Omega \Omega^T = 1$).

In this analysis we consider the case when 
the lightest active neutrino is much lighter than other 
active neutrinos:
\begin{align}
    m_3 > m_2 \gg m_1 = m_0~~\mbox{for the NO case} \,,~~~~
    m_2 > m_1 \gg m_3 = m_0~~\mbox{for the IO case} \,.
\end{align}
In addition, one of HNLs denoted by $N_S$ ($I=S$)
is assumed to have the very suppressed Yukawa coupling constants
and we take 
$\Omega_{1S} \simeq 1$ and $\Omega_{iS} \simeq 0$ ($i=2,3)$ for the NO case
and 
$\Omega_{3S} \simeq 1$ and $\Omega_{iS} \simeq 0$ ($i=1,2)$ for the IO case, 
respectively.
In this case we obtain
\begin{align}
    \label{eq:FS}
    F_S^2 \equiv (F^\dagger F)_{SS} \simeq \frac{M_S m_0}{\langle H \rangle^2}\,,
\end{align}
and the Yukawa interaction of $N_S$ becomes very suppressed
as $m_0$ becomes very small.

When the mass of $N_S$ is larger than the Higgs boson mass, 
it mainly decays into pairs of Higgs and lepton and the lifetime is estimated as
\begin{align}
    \tau_{N_S} = \frac{8 \pi}{F_S^2 \, M_S}
    \simeq \frac{8 \pi \, \langle H \rangle^2}{m_0 \, M_S^2}
    \simeq 5.0 \times 10^{-7} ~\mbox{sec}
    \left( \frac{10^{-9}~\mbox{eV}}{m_0} \right)
    \left( \frac{1~\mbox{TeV}}{M_S} \right)^2
    \,.
\end{align}
It is seen that the lifetime is rather long if 
the mass and Yukawa coupling constants of $N_S$
are both sufficiently small.
Interestingly, such a long-lived $N_S$ can dominate the energy of the 
universe and release an additional entropy by its decay.
Note that the $N_S$ decay becomes out of equilibrium if the Yukawa coupling constant
is small as
\begin{align}
    F_S^2 \lesssim \frac{M_S}{M_P} \,,
\end{align}
where $M_P$ is the (reduced) Planck mass, which corresponds to extremely small 
$m_0 < {\cal O}(10^{-5})$~eV.
\begin{figure}[t]
  \centerline{
  \includegraphics[width=9.3cm]{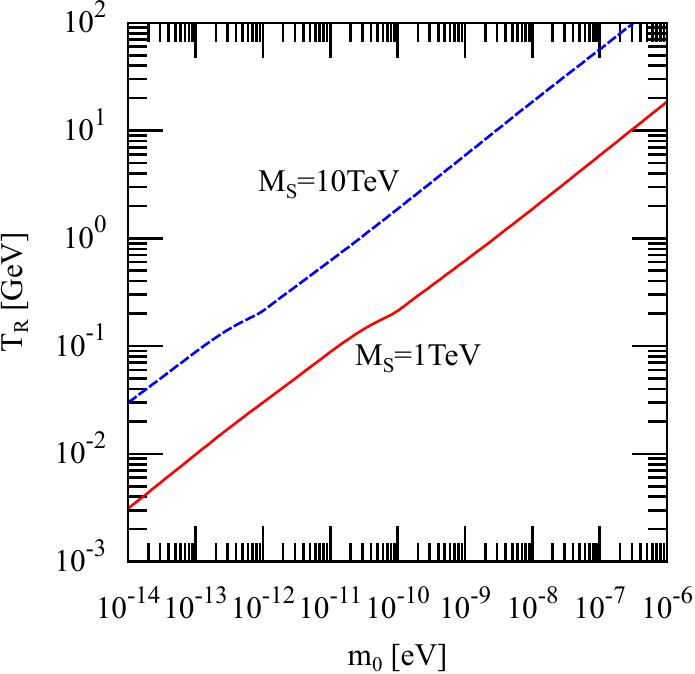}%
  \includegraphics[width=9cm]{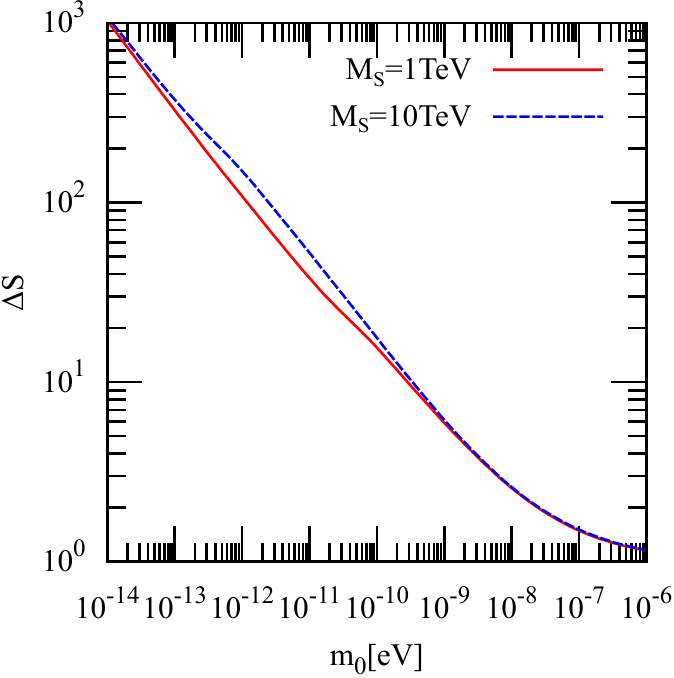}
  }%
  \vspace{-2ex}
  \caption{
    The reheating temperature $T_R$ (left) and
    the entropy production rate $\Delta S$ (right) 
    by the $N_S$ decay in terms of the lightest active neutrino mass
    $m_0$.
  }
  \label{fig:TR_DeltaS}%
\end{figure}
If this is the case, an additional entropy of the universe 
is produced and its reheating temperature
is given by
\begin{align}
    \label{eq:TR}
    T_R \sim 1~\mbox{GeV} \left( \frac{m_0}{10^{-10}~\mbox{eV}} \right)^{1/2}
    \left( \frac{M_S}{1~\mbox{TeV}} \right) \,.
\end{align}
See Fig.~\ref{fig:TR_DeltaS}.
Note that the reheating temperature is bounded from below by the cosmological 
constraints (see, for example the recent analysis in Ref.~\cite{Hasegawa:2019jsa}
and references therein).

The entropy production rate $\Delta S$ is defined by the ratio between the entropy densities
with or without the $N_S$ decay.
We estimate $\Delta S$ numerically (see, for example, Ref.~\cite{KolbTurner:1990})
and the result is also shown in Fig.~\ref{fig:TR_DeltaS}.
It is seen that $\Delta S$ is roughly given by
\begin{align}
    \label{eq:DS}
    \Delta S \sim 
    10 \left( \frac{10^{-10}~\mbox{eV}}{m_0} \right)^{1/2} \,,
\end{align}
which is almost independent on $M_S$.
 In this estimation we have not specified
the production mechanism of $N_S$ which may be related
to the inflation dynamics, 
but assumed the thermal abundance.
We find that, when the lightest active neutrino mass becomes
smaller than ${\cal O}(10^{-7})$~eV, the additional entropy can be
produced by the $N_S$ decay.  
It should be noted that 
the cosmological lower bound on $T_R$ gives the upper bound
on $\Delta S$.

\section{Gravitational wave background and neutrino masses} 
Now let us discuss the impacts of the entropy production on the
primordial GW background.
First, we briefly summarize the spectrum of the GWs.
The energy density of the GWs is given by~\cite{Maggiore:2007}
\begin{align}
    \rho_{\rm GW} 
    = \frac{1}{32 \pi G}
    \left\langle (\dot h_{ij} )^2 \right\rangle \,,
\end{align}
where $h_{ij}$ is the tensor metric perturbation which 
satisfies the transverse-traceless condition 
$\partial^i h_{ij} = h^i{}_i = 0$, and the bracket indicates
the spacial average.
The GW spectrum is expressed as
\begin{align}
    \Omega_{\rm GW} (k) \equiv 
    \frac{1}{\rho_{\rm cr}} 
    \frac{d \rho_{\rm GW}}{d \ln k}
    = \frac{1}{12}
    \left( \frac{k}{aH} \right)^2
    {\cal P}_T (k)
    \,,
\end{align}
where $\rho_{\rm cr}$ is the critical density
and ${\cal P}_T(k)$ is the tensor power spectrum expressed as
\begin{align}
    {\cal P}_T(k) = T_T^2(k) \, {\cal P}_{T}^{\rm prim} (k) \,,
\end{align}
where $T_T^2 (k)$ denotes the transfer function and
we use here the results in Ref.~\cite{Kuroyanagi:2014nba}.
The primordial tensor power spectrum 
is parameterized as
\begin{align}
    {\cal P}_T^{\rm prim} (k)
    =
    A_T (k_\ast) \left( \frac{k}{k_\ast} \right)^{n_T} \,,
\end{align}
where $A_T(k_\ast)$ and $n_T$ are the amplitude and 
the spectrum index at $k = k_\ast = 0.05$~Mpc$^{-1}$.
The amplitude is given by $A_T(k_\ast) = r \, {\cal P}_S^{\rm prim}(k_\ast)$ where the power spectrum of the scalar perturbation
is measured precisely 
as ${\cal P}_S^{\rm prim}(k_\ast) = 2.0989 \times 10^{-9}$
and the tensor-to-scalar ratio
$r$ is bounded as $r < 0.063$~\cite{Akrami:2018odb}.

The thermal history of the universe is encoded in
the transfer function.  When the entropy production 
at late time occurs by the $N_S$ decay,
the energy starts to be dominated by $N_S$ 
at some moment
and the matter dominated universe is realized
after the reheating of the primordial inflation,
and then its decay into radiations leads
to the reheating again. 
Consequently, the GW spectrum is suppressed
at frequencies higher than $f_R$ compared with 
the case without the entropy production~\cite{Seto:2003kc}.

\begin{figure}[t]
  \centerline{
  \includegraphics[width=9cm]{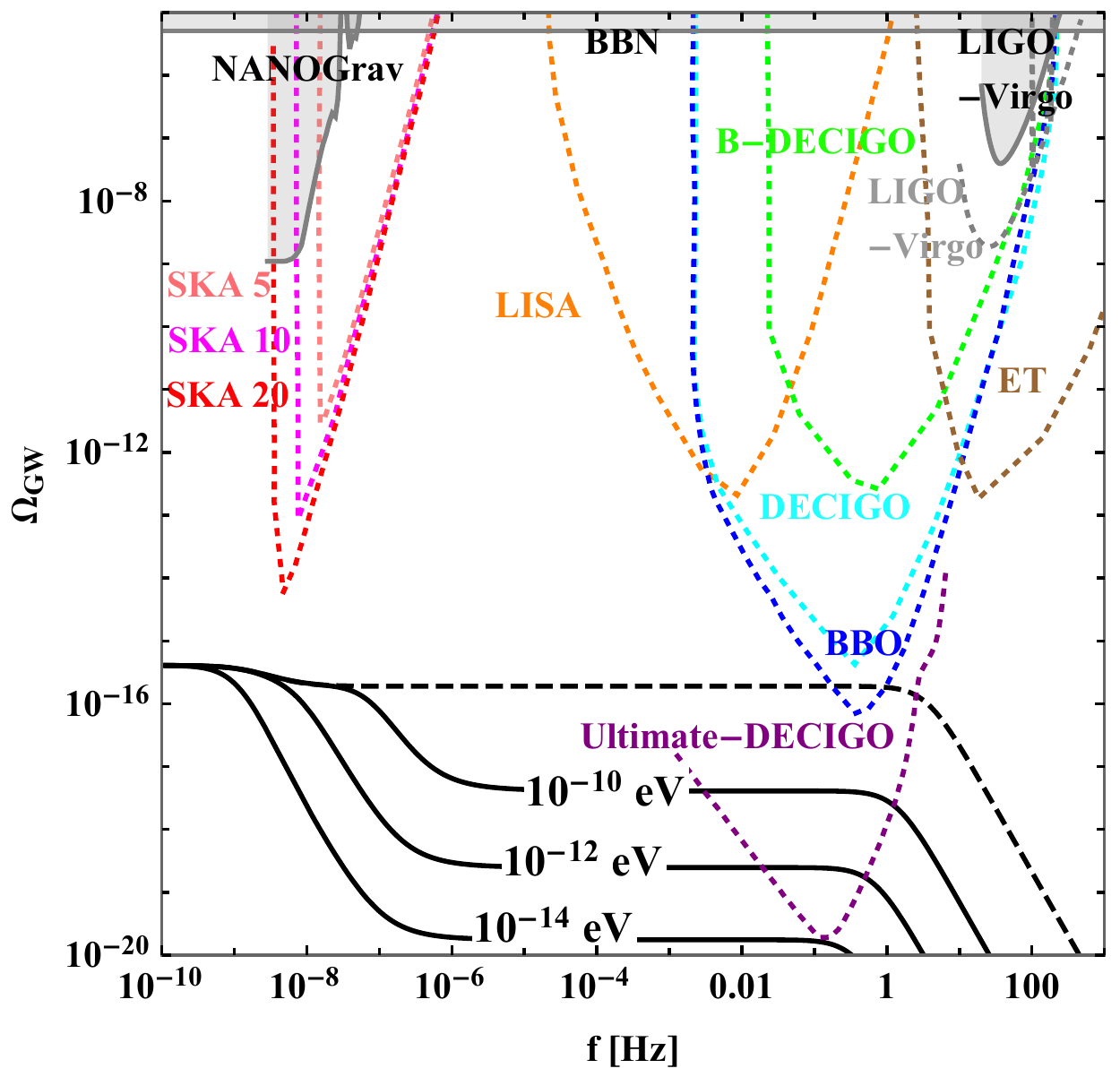}%
  \includegraphics[width=9cm]{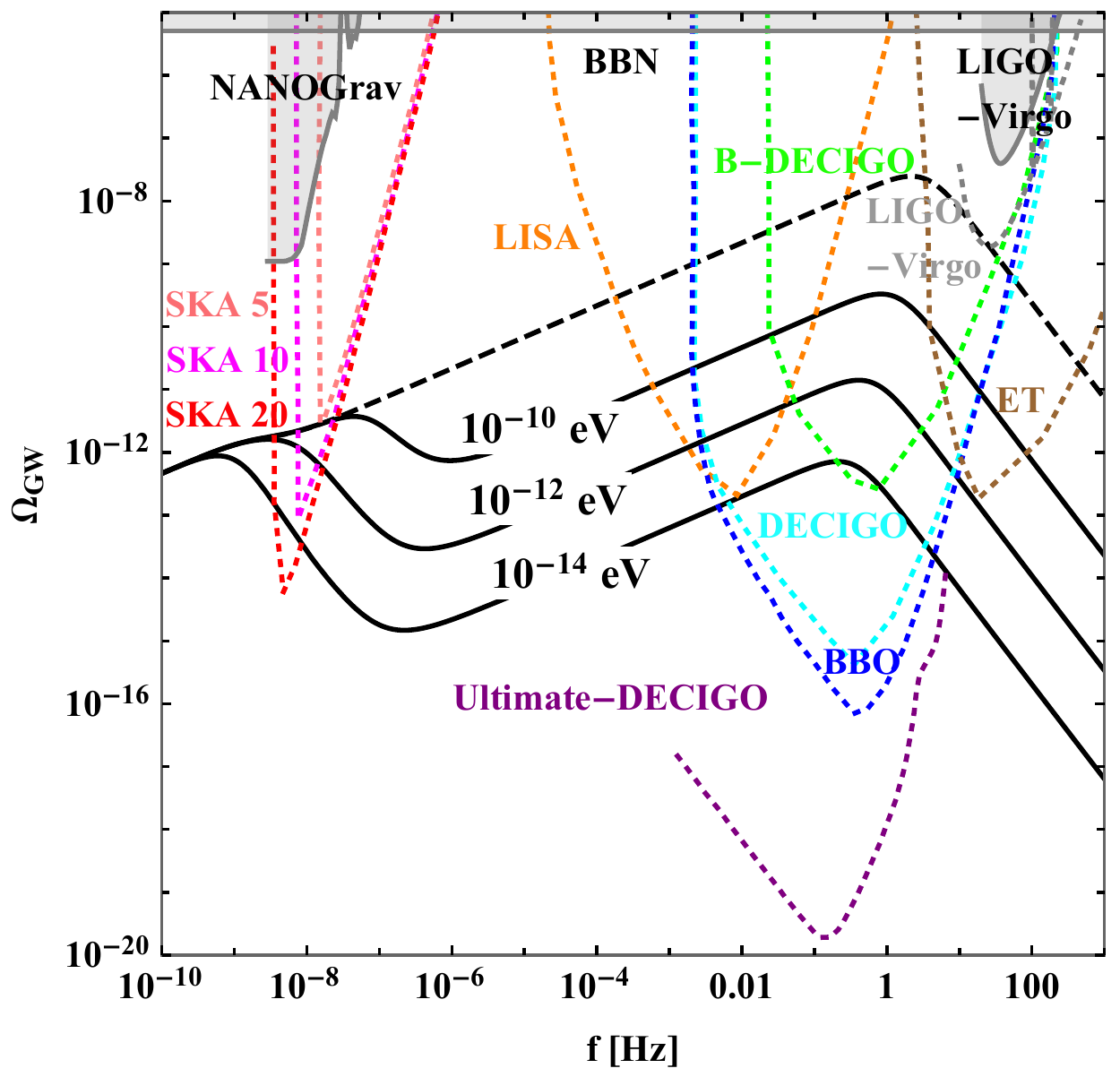}
  }%
  \vspace{-2ex}
  \caption{
    Spectra of the primordial GW background
    $\Omega_{GW}$ for the case when
    $M_S = 10$~TeV and
    $m_0 = 10^{-14}$, $10^{-12}$ and $10^{-10}$~eV by black-solid lines.
    We also show the spectrum without the entropy production by black-dashed line.
    We take $r=0.06$, $T_{RI} = 10^5$~TeV,
    and $n_T = 0$ (left) and $0.5$ (right).
    The shaded regions are excluded from BBN~\cite{Calcagni:2020tvw}, 
    LIGO-Virgo~\cite{LIGOScientific:2019vic} and NANOGrav~\cite{Arzoumanian:2018saf}.
    The dotted lines show the sensitivities by the GW observations 
     (see the details in the text). 
  }
  \label{fig:OM_GW}%
\end{figure}
\begin{figure}[t]
  \centerline{
  \includegraphics[width=9cm]{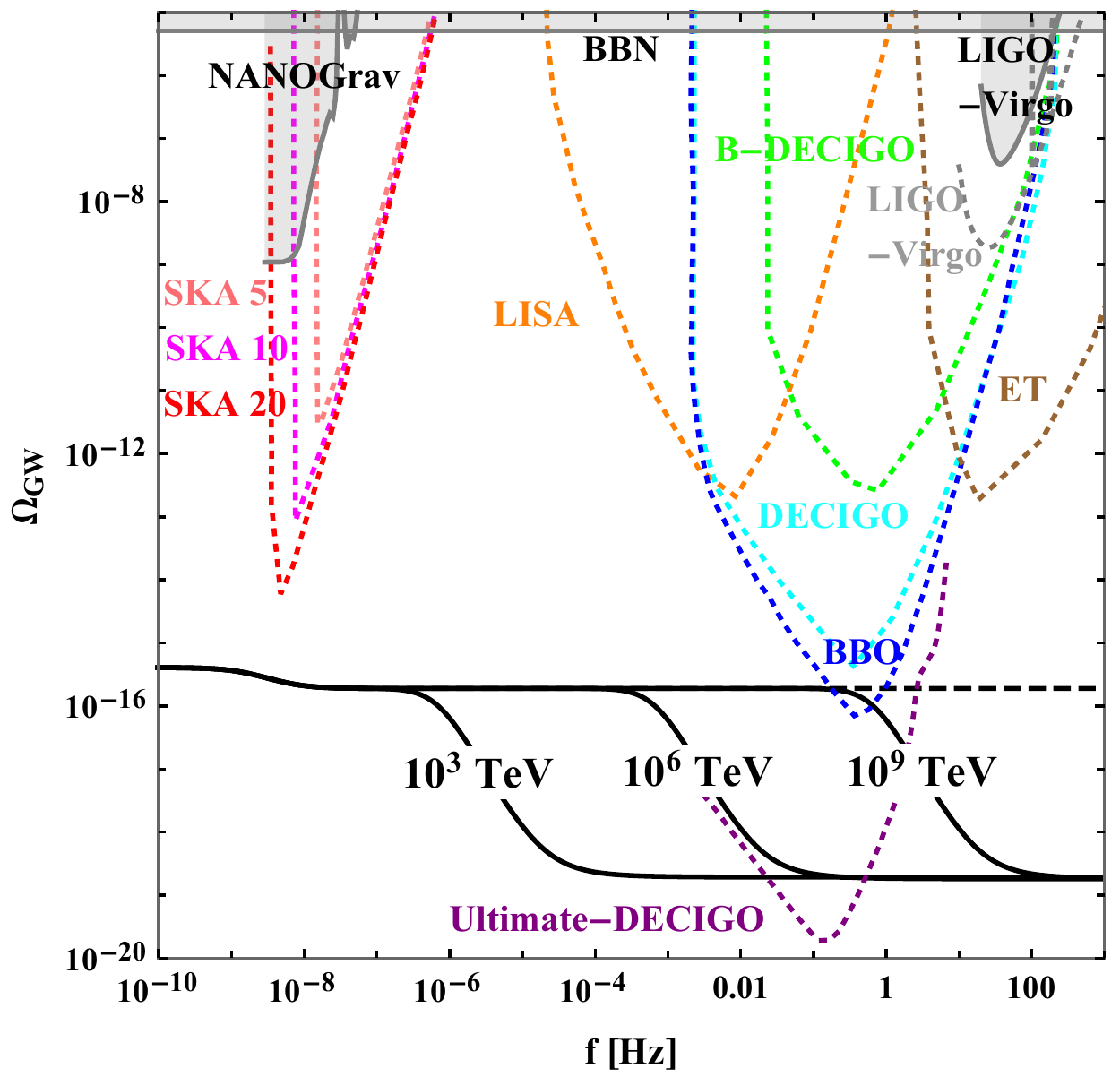}%
  \includegraphics[width=9cm]{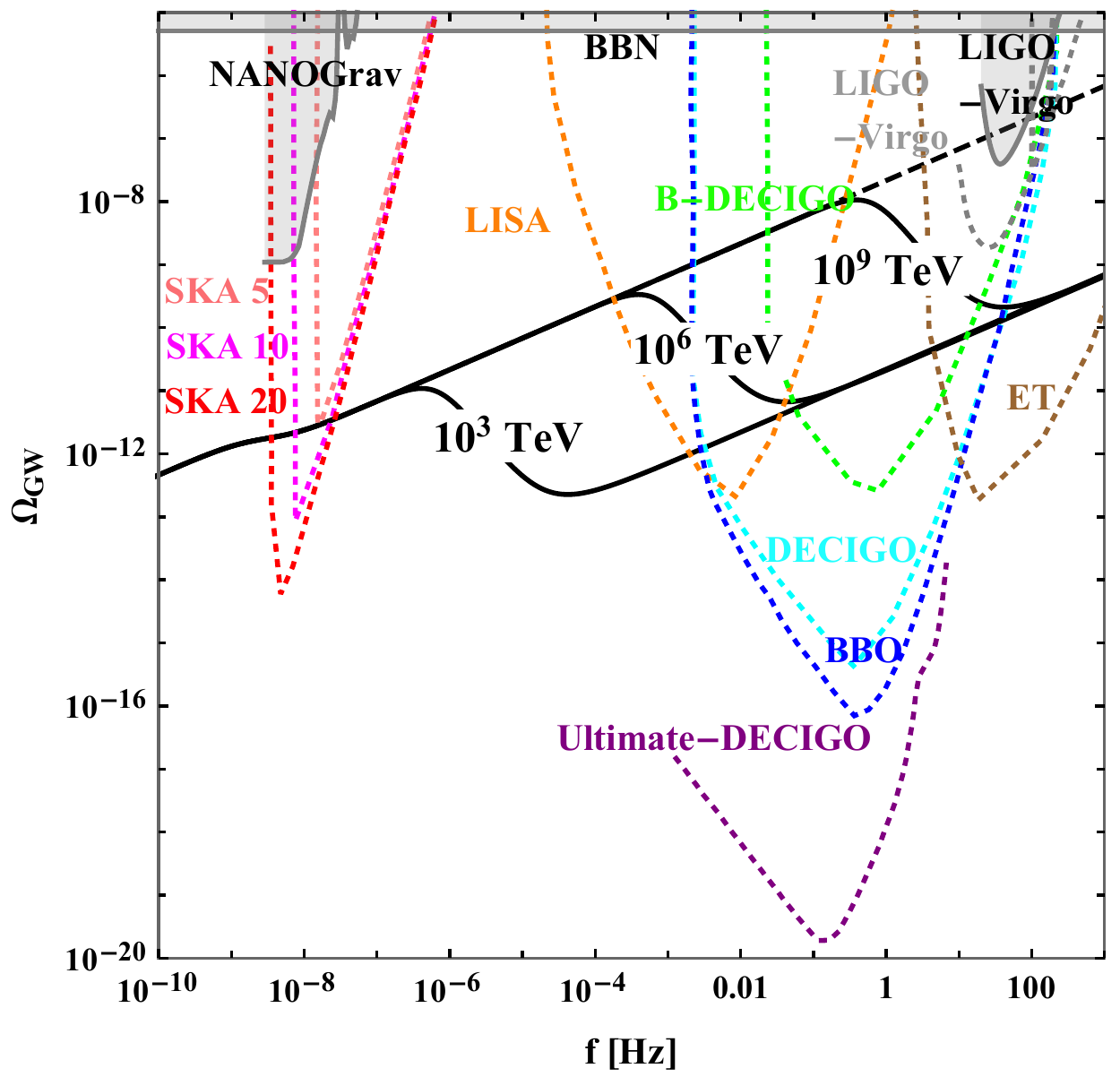}
  }%
  \vspace{-2ex}
  \caption{
    Spectra of the primordial GW background
    $\Omega_{GW}$ for the case when
    $m_0 = 10^{-12}$~eV and $M_S = 10^3$,
    $10^6$, $10^9$~TeV by black-solid lines.
    We also show the spectrum without the entropy production by black-dashed line.
    We take $r=0.06$, $T_{RI} = 10^{12}$~TeV, and $n_T = 0$ (left) and $0.5$ (right).
    See also the caption in Fig.~\ref{fig:OM_GW}.
  }
  \label{fig:OM_GW2}%
\end{figure}
In Fig.~\ref{fig:OM_GW} 
we show the spectrum $\Omega_{GW}$
for the case when $M_S = 10$~TeV and 
$m_0 = 10^{-14}$, $10^{-12}$ and 
$10^{-10}$~eV.%
\footnote{To make a precise estimation of the GW spectrum we have to take into account the damping effect to the GW spectrum due to the free-streaming of $N_S$, which we have neglected in this analysis
since it is expected to be small. Our final results do not change much by this effect.
}
Here we take $r=0.06$, $T_{RI} =10^{5}$~TeV
(the reheating temperature of the primordial inflation), 
and $n_T = 0$ and $0.5$.
For reference we also present the result without the entropy production.
It is seen that $\Omega_{GW}$ is suppressed for the higher frequencies
$f \gtrsim f_R$,  where the critical frequency is given by
\begin{align}
    f_R \sim
    10^{-11}~\mbox{Hz} \,
    \left( \frac{T_R}{10~\mbox{MeV}} \right) \,.
\end{align}
Note that $f_R$ is sensitive to $T_R$, and hence to $m_0$ and $M_S$
as shown in Eq.~(\ref{eq:TR}).
On the other hand, the magnitude of the spectrum suppression is expressed as
\begin{align}
    \Delta \Omega_{GW} = 
    \frac{\left. \Omega_{GW} \right|_{\rm w EP} }
    {\left.\Omega_{GW} \right|_{\rm wo EP}  }
\end{align}
where $\left .\Omega_{GW} \right|_{\rm w EP}$
and $\left. \Omega_{GW} \right|_{\rm wo EP}$ are
the GW spectrum for $f \gg f_R$ with and without the entropy production,
respectively.  This suppression factor has been estimated as~\cite{Seto:2003kc}
\begin{align}
    \Delta \Omega_{\rm GW}
    \simeq \frac{1}{\Delta S^{4/3}} \,.
\end{align}
It is then found from Eq.~(\ref{eq:DS}) that
$\Delta \Omega_{\rm GW}$ gives the information of $m_0$.

\begin{figure}[t]
    \vspace{5cm}
  \centerline{
\includegraphics[width=9cm]{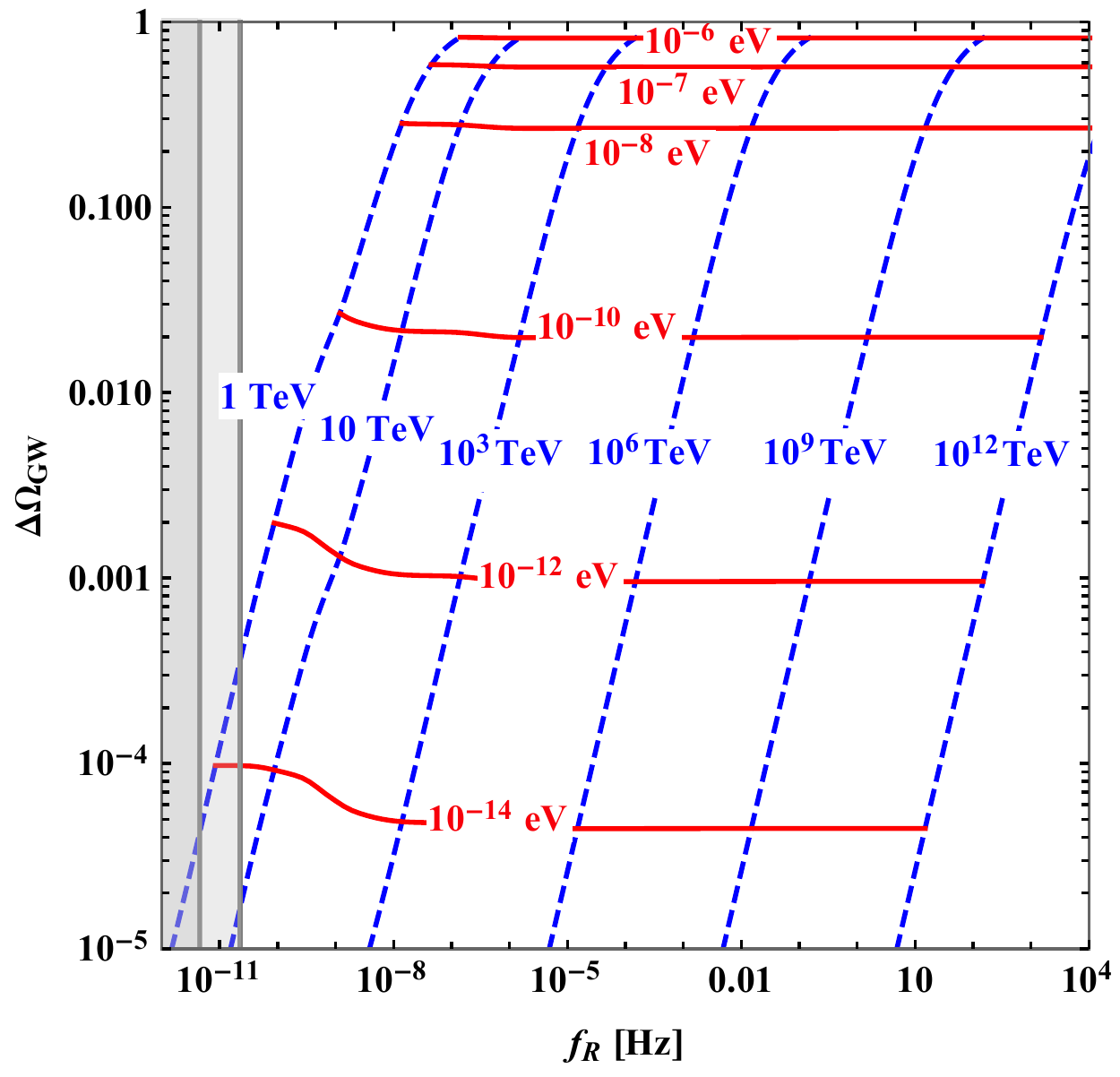}
  }%
  \vspace{-2ex}
  \caption{
  The indicated values of the masses
  of the lightest active neutrino $m_0$ (red-solid lines)
  and the right-handed neutrino $M_S$ (blue-dashed lines)
  in terms of the critical frequency $f_R$ and the suppression factor
  $\Delta \Omega_{GW}$. Gray-shaded regions are excluded by 
  the BBN bounds ($T_R \ge 1$ and 5~MeV).
  }
  \label{fig:m0MS}%
\end{figure}

In Fig.~\ref{fig:OM_GW} we also show the upper bounds on $\Omega_{GW}$ from  BBN~\cite{Boyle:2007zx,Calcagni:2020tvw}, 
LIGO-Virgo~\cite{LIGOScientific:2019vic} and NANOGrav~\cite{Arzoumanian:2018saf}.
In addition, we show the sensitivities by the future GW observations:
SKA~\cite{Janssen:2014dka},
LISA~\cite{Audley:2017drz},
ET~\cite{Sathyaprakash:2012jk}, BBO~\cite{Crowder:2005nr},
(B-)DECIGO~\cite{Seto:2001qf,Sato:2017dkf} 
and Ultimate-DECIGO~\cite{Kudoh:2005as}.
It is found that,
when the mass of $N_S$ is ${\cal O}(10)$~TeV
and $n_T$ is a relatively large value,%
\footnote{
The possible models realizing such a large value of $n_T$
have been proposed in the context of modified gravity and non-standard inflation models.  See, for example, Ref.~\cite{Kuroyanagi:2020sfw}
and references therein.}
the predicted $f_R$ can be probed by the 
pulsar time array observations for $m_0 ={\cal O}(10^{-14})$--${\cal O}(10^{-10})$~eV, and $\Delta \Omega_{GW}$ can be probed by the GW interferometers.
On the other hand, we show in Fig.~\ref{fig:OM_GW2}
the GW spectrum $\Omega_{GW}$ with $M_S =10^3$, $10^6$ and $10^9$~TeV
by taking $m_0 =10^{-12}$ eV.  It is found that
the effect by $N_S$ with masses $M_S > {\cal O}(10^6)$~TeV
can be probed by the future GW observations if $n_T$ is a relatively large.

As shown above, the distortion of the GW spectrum 
due to the entropy production by $N_S$
can be probed by the future observations.
Importantly, we can reconstruct the masses of the lightest active neutrino 
$m_0$ and the right-handed neutrino $N_S$ if $f_R$ and
$\Delta \Omega_{GW}$ will be provided by the observations.
It should be noted that the Yukawa coupling $F_S$ 
can be determined from $m_0$ and $M_S$ as shown in Eq.~(\ref{eq:FS}).
This point is represented in Fig.~\ref{fig:m0MS},
where we present the indicated values of $m_0$ and $M_S$
for given $f_R$ and $\Delta \Omega_{GW}$.
The result for the range $M_S = 1$~TeV to $10^{12}$~TeV
is shown.
We find that the mass of the lightest active neutrino
with $m_0 < {\cal O}(10^{-7})$~eV can be probed which 
is very difficult to examine by the neutrino experiments.

Before closing this section, we mention the mass range of the 
right-handed neutrino $N_S$.  We have considered the case
when $M_S > {\cal O}(1)$~TeV so far.
The extension to the lighter mass region can be done
in a straightforward way by taking into account the appropriate
decay modes of $N_S$.  
This issue will be discussed elsewhere~\cite{AO}.

\section{Conclusions} 
We have considered the Standard Model with three right-handed neutrinos
which realizes the seesaw mechanism for the observed tiny neutrino masses.
Especially, we have investigated the case that
one of three right-handed neutrinos, $N_S$, have very suppressed
Yukawa coupling $F_S$, and the lightest neutrino mass $m_0$
becomes smaller than ${\cal O}(10^{-7})$~eV.  
In this case the late-time entropy production occurs by the $N_S$ decay
and can modify the spectrum of the primordial gravitational wave background
significantly.  The spectrum can be suppressed for the frequencies
$f > f_R$ by the factor $\Delta \Omega_{GW}$.
We have shown that the observational data of $f_R$ and $\Delta \Omega_{GW}$
determines both the mass of the lightest active neutrino$m_0$ and the $N_S$ mass $M_S$, 
which leads to the determination of the Yukawa coupling $F_S$.  
It has been found that
the very small value of $m_0 < {\cal O}(10^{-7})$~eV,
which is very difficult to test by the neutrino experiments,
can be probed by the GW spectrum shape by the future gravitational wave
detection projects and the pulsar timing arrays.

\section*{Acknowledgments}
The work of T.A. was partially supported by JSPS KAKENHI
Grant Numbers 17K05410, 18H03708, 19H05097, and 20H01898.
This work of H.O. was partially supported by the Sasakawa Scientific Research Grant from The Japan Science Society (Grant Number 2019-2022).


\end{document}